\documentclass[11pt]{article}

\usepackage[utf8]{inputenc}
\usepackage[T1]{fontenc}
\usepackage{lmodern}
\usepackage{graphicx}
\usepackage{amsmath}
\usepackage{amssymb}
\usepackage{booktabs}
\usepackage{hyperref}
\usepackage{multirow}
\usepackage{float}
\usepackage{enumitem}
\usepackage{url}
\usepackage{color}
\usepackage{listings}
\usepackage{tikz}
\usetikzlibrary{positioning,arrows.meta}

\title{
TeleTwin: A Calibrated Digital Twin of Telegram Crisis Discourse as a Ground-Truth Testbed for Detecting LLM-Powered Information Operations
}

\author{
Despoina Antonakaki \\
Institute of Computer Science, FORTH, Heraklion, Greece \\
Technical University of Crete, Chania, Greece \\
\texttt{despoina@ics.forth.gr}, \texttt{dantonakaki@tuc.gr}
\and
Sotiris Ioannidis \\
Technical University of Crete, Chania, Greece \\
\texttt{sioannidis@tuc.gr}
}

\date{2026}

\begin{document}

\maketitle

\begin{abstract}
Large language models make it cheap to run influence operations whose messages
never repeat, yet real platforms offer almost no ground truth for testing
whether coordination detectors can find them. The problem is sharpest on
Telegram, which moderates little and spreads content through broadcast channels
and forwarding rather than a follow graph. We introduce TeleTwin, an LLM-powered
digital twin of a Telegram crisis-discourse ecosystem in which channels are
agents, each instantiated from the history of a real channel, calibrated on 1.07
million messages from 195 public channels on the Israel--Gaza and Iran--Israel
conflict (March--August 2025). On a held-out period containing the June 2025
Iran--Israel war, the twin tracks daily dynamics better than statistical
baselines (correlation 0.86 against 0.64) and reacts to the war unprompted, but
it over-focuses on the conflict, over-amplifies forwarding cascades,
underestimates the shock and writes recognisably LLM-like text. We then inject
LLM-written campaigns with known membership into the real data and into the
twin. In real data, identical-text, co-URL and synchrony detectors fail once a
campaign rewords, staggers or drops links; multilingual semantic similarity
recovers loud campaigns, including cross-lingual ones, but not quiet ones, and
its false positives fall unevenly across political perspectives. Inside the
twin all detectors fail, because the simulated background is about 40 times
more semantically homogeneous than the real one. Digital twins can supply
ground truth for campaigns, but not, without calibrating their diversity, a
realistic background for evaluating semantic detectors.
\end{abstract}

\section{Introduction}

Large language models (LLMs) have lowered the cost of producing fluent,
persona-consistent political text to nearly zero. Analysts have warned that
this changes the economics of influence operations: campaigns that once needed
teams of writers can now be run by a handful of operators who generate
varied, on-message content at scale~\cite{goldstein2023generative,bontcheva2024generative}.
Automated manipulation of social media is not new: spam campaigns that hijack
trending topics and networks of social bots have been documented for over a
decade~\cite{antonakaki2021survey,antonakaki2016spam,ferrara2016rise}. What is
new is the quality of the text. LLM-driven accounts are no longer hypothetical: networks of automated accounts
posting LLM-generated text have already been observed in the
wild~\cite{yang2024anatomy}. For defenders, this poses a measurement problem as
much as a detection problem. Coordination detectors have mostly been developed
and validated on campaigns that reuse identical links, hashtags or
text~\cite{pacheco2021uncovering,giglietto2020takes,nizzoli2021coordinated}, and
it is not known how they behave when every message in a campaign is a fresh
paraphrase, or a translation into another language.

Answering that question on real platforms is hard because there is almost never
ground truth. We rarely know with certainty which channels in a crisis-time
information ecosystem are coordinated, or which messages were machine-generated.
Evaluations therefore rely on proxy labels such as platform takedowns or
suspensions~\cite{shevtsov2023russiaukraine}, which are incomplete and biased
towards what platforms already detect. This gap is sharpest on Telegram. The
platform is central to wartime information flows around Russia--Ukraine and the
Middle East~\cite{bawa2025telegram,antonakaki2025israelhamaswar,antonakaki2026iran},
yet it moderates little, publishes no takedown data, and exposes a structure
unlike that of microblogs: content is broadcast by channels to passive
subscribers and spreads through forwarding chains, not through a follow graph
and a like-driven feed.

LLM-powered social simulations offer a way to create ground truth
by construction. Generative agents can reproduce believable individual and
collective behaviour~\cite{park2023generative,park2022social}, and recent systems
simulate whole social platforms with LLM agents~\cite{gao2023s3,tornberg2023simulating,mou2024unveiling}.
Y Social~\cite{rossetti2024ysocial} goes furthest, framing such a system as a
\emph{digital twin} of an online social platform. However, these systems share
two limitations for our purpose. First, they model microblog or forum
platforms: follows, likes, feed recommenders and comment threads. None of these
mechanics describe Telegram. Second, they are mostly demonstrated with
descriptive case studies rather than tested against held-out real data, so it
is unclear how far their outputs can be trusted as a stand-in for a real
ecosystem.

In this paper we introduce \textbf{TeleTwin}, a digital twin of a Telegram
crisis-discourse ecosystem, and use it as a ground-truth testbed for detecting
LLM-powered information operations. TeleTwin treats \emph{channels} rather
than individual users as agents, reflecting Telegram's broadcast structure: each
simulated channel is instantiated from the posting history of a real channel,
including its language, topics, activity rhythm and the channels it forwards
from. We calibrate it on a corpus we collected from public Telegram channels
covering the Israel--Gaza and Iran--Israel conflict between March and August
2025, discovered by following real forwarding links and balanced with a
symmetric search for under-represented perspectives. The period contains a
sharp exogenous shock, the June 2025 Iran--Israel war, which we hold out from
calibration and use to test whether the twin responds to real events the way
the real ecosystem did. The corpus contains over one million messages from 195
channels, 215{,}722 of them about the conflict.

We address two research questions:
\begin{itemize}[leftmargin=*]
  \item \textbf{RQ1 (Fidelity).} How closely does a calibrated LLM-based twin
    reproduce the topical, temporal and forwarding dynamics of a real Telegram
    ecosystem on a held-out period containing an exogenous shock, and where
    does it fail?
  \item \textbf{RQ2 (Detection).} How well do coordination detectors identify
    LLM-powered campaigns whose ground truth is known, as the campaigns vary in
    style (copy-paste, paraphrase, cross-lingual translation), timing, scale,
    and embedding in organic forwarding?
\end{itemize}

Our contributions are:
\begin{enumerate}[leftmargin=*]
  \item \textbf{A Telegram-native digital twin.} To our knowledge, TeleTwin is
    the first LLM-based simulator built around Telegram's primitives
    (channels, forwards, views), with agents instantiated from real channels.
  \item \textbf{A held-out-shock fidelity protocol.} We evaluate the twin on a
    period it never saw, across a real crisis, against a non-LLM baseline,
    and report failure modes as well as matches.
  \item \textbf{A ground-truth benchmark.} We inject LLM-powered campaigns into
    both real data and the twin, producing labelled Telegram data in which
    campaign membership is known exactly.
  \item \textbf{A detector evaluation, including fairness across perspectives.}
    We measure how lexical, co-sharing and semantic coordination detectors
    degrade under LLM paraphrase and translation, and report false-positive
    rates separately for each political perspective in the ecosystem.
\end{enumerate}
Code, data (without message texts) and all results are publicly available on
Zenodo~\cite{antonakaki2026teletwinzenodo} at
\url{https://zenodo.org/records/23187748}.

\section{Related Work}

\subsection{LLM-Based Social Simulation and Digital Twins}
Agent-based models have long been used to study opinion dynamics, polarisation
and information diffusion on online platforms, typically with simple
rule-following agents whose behaviour is sampled from fixed
distributions~\cite{pappalardo2024survey}. Generative agents replaced these
rules with LLMs: Park et al.\ showed that LLM agents with memory produce
believable individual and group behaviour in a sandbox
environment~\cite{park2023generative}, and earlier populated prototypes of
online communities with LLM-generated
participants~\cite{park2022social}. Subsequent work scaled this to
social networks. S$^3$ simulates information, emotion and attitude propagation
with LLM-empowered agents~\cite{gao2023s3}; T\"ornberg et al.\ use LLM agents on
a simulated platform to compare news-feed algorithms~\cite{tornberg2023simulating};
Mou et al.\ simulate large-scale social movements on a Twitter-like
environment~\cite{mou2024unveiling}; and Chuang et al.\ reproduce opinion
dynamics with networks of LLM agents~\cite{chuang2023simulating}. Populations of
LLM agents have also been shown to form scale-free networks when left to choose
their own connections~\cite{demarzo2023emergence}. Closest to our goals,
LLM-agent simulations have recently been used to study misinformation
propagation~\cite{maurya2025simulating} and bot-driven manipulation of beliefs
in agent networks~\cite{burmester2026belief}.

Y Social~\cite{rossetti2024ysocial} frames an LLM-powered platform simulator as
a digital twin, in the engineering sense of a virtual replica that supports
experimentation on its physical counterpart~\cite{barricelli2019survey}. It
exposes microblog primitives (post, comment, react, follow, share news), assigns
agents personas with political leanings and Big Five traits, and supports
pluggable content and follow recommenders. Its case study is descriptive: it
shows that agents produce plausible threads and heavy-tailed activity, but does
not compare the simulation against real data from the platform it replicates.
TeleTwin differs in three ways: it models Telegram's broadcast-and-forward
structure instead of a follow graph and feed; its agents are instantiated from
specific real channels rather than sampled personas; and it is evaluated against
held-out real data spanning an exogenous shock.

\subsection{Coordinated Inauthentic Behaviour Detection}
Early work on automated manipulation focused on detecting individual social
bots from account features and behaviour~\cite{ferrara2016rise,cresci2020decade,antonakaki2021survey},
including explainable and generic machine-learning detectors trained on
election-time and platform-suspension data~\cite{shevtsov2022icwsm,shevtsov2022bots,shevtsov2025botartist},
and showed that bots can shape exposure to inflammatory
content~\cite{stella2018bots}. Because sophisticated operations mix automated
and human-run accounts, attention shifted from individual accounts to
\emph{coordination} between them. Pacheco et al.\ propose a general framework
that builds networks of accounts linked by unexpectedly similar behaviour, such
as shared content, hashtags or near-simultaneous activity~\cite{pacheco2021uncovering}.
Giglietto et al.\ detect coordinated link sharing, i.e.\ groups of pages that
post the same URLs within short time windows~\cite{giglietto2020takes}, and
Nizzoli et al.\ characterise coordinated communities during the 2019 UK general
election~\cite{nizzoli2021coordinated}. Earlier, spam campaigns were identified
through their coordinated abuse of trending topics, where many accounts push the
same content into a trend at the same time~\cite{antonakaki2015thinkrt,antonakaki2016spam};
these are exactly the kind of surface signals that LLMs now make easy to avoid. Recent work extends detection across
platforms, including Telegram, in the run-up to the 2024 US
election~\cite{cinus2025crossplatform}. In our own prior work we analysed
coordinated dissemination across Telegram and Reddit during political
turbulence~\cite{antonakaki2026venezuela}, and combined lexical diversity,
temporal burstiness, rhetorical repetition and semantic homogenisation into a
synthetic narrative coordination score~\cite{antonakaki2026synthetic}.

Most of these methods rely on surface overlap: identical URLs, repeated
hashtags, copied text, or tight temporal synchrony. LLMs remove much of that
overlap at no cost, since each message in a campaign can be a distinct
paraphrase or a translation into another language. LLM-powered botnets have
been documented in the wild~\cite{yang2024anatomy}, and BotSim showed that bot
detectors trained on traditional datasets lose accuracy on a simulated
LLM-driven botnet~\cite{qiao2025botsim}. Machine-generated-text detectors such as
DetectGPT~\cite{mitchell2023detectgpt} offer a complementary, content-level
signal, but they work on individual messages rather than coordination, and
their reliability on short social-media text in several languages is limited.
Semantic similarity over multilingual sentence embeddings~\cite{reimers2019sentencebert}
is a natural candidate for recovering coordination that survives paraphrase and
translation; how much it recovers is one of the questions our benchmark is
designed to answer. BotSim is the closest prior work in spirit, but it targets
account-level bot detection on a Twitter-like platform, whereas we target
channel-level coordination on Telegram.

\subsection{Information Operations on Telegram}
Telegram combines public broadcast channels, large groups and near-absent
moderation, which has made it a key venue for alternative and extremist
communities and for state-aligned messaging. The Pushshift Telegram dataset
provided one of the first large-scale public collections of Telegram
channels~\cite{baumgartner2020pushshift}, and Urman and Katz used forwarding
links to map far-right channel networks, showing that forwards expose the
structure of Telegram communities~\cite{urman2022what}. During the
Russia--Ukraine war, Kremlin-related channels on Telegram have been shown to
act as a coordinated communication infrastructure~\cite{bawa2025telegram}. Our
previous studies analysed the Israel--Hamas war, the Iran--Israel escalation and
the Venezuelan crisis across Telegram, Reddit and
Twitter~\cite{antonakaki2025israelhamaswar,antonakaki2025crossplatformisrael,antonakaki2026iran,antonakaki2026venezuela},
and compared pro-Israel and pro-Palestine discourse on Telegram through
sentiment, stance and framing~\cite{zafeiropoulos2026telegram}. This work builds
on a longer line of studies of social media during political turbulence and
crises, from the 2015 Greek referendum and elections~\cite{antonakaki2016referendum,antonakaki2017political}
and cross-platform analysis of the 2020 US elections on Twitter and
YouTube~\cite{shevtsov2023youtube} to large-scale datasets of the
Russo-Ukrainian war~\cite{shevtsov2022ukrainedataset,lamprou2025crisis}.
All of this work is observational: it describes real channels but cannot say
with certainty which of them are coordinated or how well a detector would find
a campaign that was actually there. TeleTwin complements it by providing a
controlled setting in which campaigns are injected with known ground truth,
calibrated on the same kind of crisis-time Telegram ecosystem these studies
describe.

\section{Data: A Telegram Crisis-Discourse Ecosystem}
\label{sec:data}

TeleTwin is calibrated and evaluated on a corpus of public Telegram channels
that we collected for this study. The unit of analysis is the \emph{channel}: a
one-to-many broadcast feed whose messages are pushed to subscribers and can be
forwarded, with attribution, into other channels. The corpus covers the
Israel--Gaza and Iran--Israel conflict between 1 March and 31 August 2025.
Unlike keyword-based collections, we retain \emph{every} message each channel
published in this window, because the twin must reproduce whole-channel
behaviour (posting rhythm, off-topic content, forwarding habits), not only
on-topic text. All collection used the official Telegram API through the
Telethon client, and was restricted to public broadcast channels.

\subsection{Building the Sampling Frame}
Channels entered the corpus in three stages, summarised in
Table~\ref{tab:frame}. Every channel was resolved against the live platform and
checked for subscriber count and activity; no channel handle was entered by
hand except the six seeds, which were verified in our earlier
work~\cite{antonakaki2025israelhamaswar,antonakaki2026iran}.

\paragraph{Stage 1: forward-graph snowball.}
\begin{sloppypar}
Starting from six seed channels spanning pro-Western, pro-Russian, regional and
neutral outlets (\texttt{MiddleEastEye\_TG}, \texttt{DDGeopolitics},
\texttt{IntelSlava}, \texttt{rybar}, \texttt{KyivIndependent\_official},
\texttt{wartranslated}), we followed forwarding links for two hops. For each
visited channel we sampled its 300 most recent messages, measured the share
matching the domain keywords (Section~\ref{sec:tagging}), recorded the channels
it forwards from, and expanded the 15 most-forwarded sources of every channel
whose keyword share was at least 5\%. Forward links reveal which channels
actually feed each other, so the frame follows the real circulation of content
rather than our assumptions about it~\cite{urman2022what}. The snowball visited
136 channels, of which 126 met the relevance threshold.
\end{sloppypar}

\paragraph{Stage 2: forward-source expansion.}
After collecting the Stage~1 channels, we found that only a third of their
forwards originated inside the frame. Since the twin must simulate forwarding
chains, we added the 50 channels outside the frame that were forwarded most
often, regardless of topic. Messages forwarded from channels that remain outside
the frame are treated as exogenous input to the simulation, in the same way as
news headlines (Section~\ref{sec:twin}).

\paragraph{Stage 3: perspective balancing.}
\begin{sloppypar}
Only three of the six seeds forward other channels, and all three belong to the
pro-Russian or alternative-geopolitics sphere; the mainstream outlets among the
seeds forward nothing. The snowball therefore over-samples that sphere. To
correct this we ran Telegram's public channel search with five query groups of
five terms each (neutral, Israeli, Palestinian/Arab, Western media, Iranian),
using the same number of terms per group. The pro-Russian perspective was
deliberately excluded from the search because it was already over-represented.
Candidates were retained only if they (i) had at least 5{,}000 subscribers,
(ii) published at least 30 messages inside the collection window, (iii) had at
least 10\% of those messages matching the domain keywords, and (iv) were not
flagged by Telegram as scam or fake. The search returned 164 broadcast channels,
of which 30 passed. We excluded two further channels that presented themselves
as international news outlets but were unverified and appeared to be unofficial
mirrors, leaving 28 channels, including the official channels of the Israel
Defense Forces, \emph{The Times of Israel}, \emph{The Jerusalem Post}, Al
Jazeera English, Quds News Network and Press TV.
\end{sloppypar}

\subsection{Collected Fields}
For every message we store its text, timestamp, edit timestamp, view count,
forward count, reply count, total reactions, media type and album identifier.
For forwarded messages we additionally store the identifier of the source
channel, the identifier of the original message and its original timestamp.
These three fields allow each forward to be linked to the exact post it copies,
so that forwarding cascades can be reconstructed across the frame. We store no
information about subscribers or other individual users; all data concern
public channels that broadcast to an open audience.

\subsection{Domain Tagging}
\label{sec:tagging}
Messages are tagged as belonging to the Gaza/Iran domain with an English and
Russian keyword list covering the actors, places and events of the conflict
(e.g., \emph{Gaza}, \emph{Hamas}, \emph{IDF}, \emph{Iran}, \emph{IRGC},
\emph{Netanyahu}, and their Russian equivalents), derived from the event
definitions of our earlier studies. English keywords are matched on word
boundaries, so that short terms such as \emph{IDF} do not match inside unrelated
words; Russian keywords are matched as stems so that inflected forms are
counted. Tags are used for analysis and evaluation only: they do not filter what
is collected.

\subsection{Calibration and Held-Out Periods}
We split the window at 1 June 2025. The \emph{calibration period} (March--May)
is the only data the twin sees when agents are built and parameters are
fitted. The \emph{held-out period} (June--August) is used only to evaluate
fidelity (RQ1). The held-out period contains the Iran--Israel war of 13--24 June
2025, the largest exogenous shock in the window; the calibration period contains
a smaller one, the collapse of the Gaza ceasefire on 18 March 2025. Both appear
clearly in the data without being given to the tagger:
the busiest calibration-period day is 18 March (2{,}122 Gaza/Iran messages, 2.5
times the period median), and the busiest day of the whole window is 13 June,
the first day of the war (10{,}201 messages, ten times the held-out median).

\subsection{Corpus Statistics}
Table~\ref{tab:frame} summarises the corpus by how channels entered the
frame, and Table~\ref{tab:periods} gives the ecosystem-level statistics that
serve as calibration and evaluation targets. Of the 204 channels in the frame,
195 published at least one message in the window, for a total of 1{,}074{,}018
messages, of which 215{,}722 (20.1\%) are tagged Gaza/Iran (173{,}888 in
English and 41{,}834 in Russian). The balancing stage matters beyond its size:
its 28 channels contribute 109{,}068 Gaza/Iran messages, half of the domain
total, because they post almost exclusively about the conflict, whereas
channels found by the snowball cover a broad range of geopolitical topics.
Balancing channels also forward much less (2.7\% of messages versus about 12\%
elsewhere), consistent with their role as primary news sources rather than
aggregators. About 48\% of all forwards originate from channels inside the
frame; the rest enter the simulation as exogenous input.

\begin{table}[t]
\centering
\small
\caption{Corpus composition by sampling stage (1 March -- 31 August 2025).
\emph{Fwd} is the share of messages that are forwards; \emph{In-frame} is the
share of those forwards whose source channel is in the frame; \emph{Media} is
the share of messages without text. Language is assigned per text message by
majority script.}
\label{tab:frame}
\setlength{\tabcolsep}{4pt}
\resizebox{\textwidth}{!}{%
\begin{tabular}{lrrrrrrrr}
\toprule
Stage & Channels & Messages & Gaza/Iran & Fwd & In-frame & Media & English & Russian \\
\midrule
Snowball          & 117 &   527{,}678 &  62{,}137 & 11.7\% & 57.8\% & 22.1\% & 195{,}007 & 216{,}033 \\
Forward sources   &  50 &   351{,}929 &  44{,}517 & 12.3\% & 35.8\% & 24.9\% & 130{,}071 & 134{,}239 \\
Balancing         &  28 &   194{,}411 & 109{,}068 &  2.7\% & 23.7\% & 17.8\% & 139{,}829 &  19{,}940 \\
\midrule
Total             & 195 & 1{,}074{,}018 & 215{,}722 & 10.3\% & 47.6\% & 22.2\% & 464{,}907 & 370{,}212 \\
\bottomrule
\end{tabular}}
\end{table}

\begin{table}[t]
\centering
\small
\caption{Ecosystem-level statistics for the calibration and held-out periods.
A cascade is the set of in-frame channels that forwarded the same original
post; forward delay is the time between the original post and its forward.}
\label{tab:periods}
\begin{tabular}{lrr}
\toprule
 & \begin{tabular}[b]{@{}r@{}}Calibration\\(Mar--May)\end{tabular} & \begin{tabular}[b]{@{}r@{}}Held-out\\(Jun--Aug)\end{tabular} \\
\midrule
Active channels                         & 183 & 195 \\
Messages                                & 508{,}353 & 565{,}665 \\
Gaza/Iran msgs per day (median)          & 856 & 1{,}003 \\
Gaza/Iran msgs, busiest day              & 2{,}122 (18 Mar) & 10{,}201 (13 Jun) \\
Forward edges inside the frame          & 1{,}038 & 1{,}152 \\
Forwards from inside the frame           & 46\% & 49\% \\
Cascades of size $\geq 2$               & 3{,}620 & 3{,}531 \\
Cascades of size $\geq 4$               & 342 & 550 \\
Forward delay, 10th percentile          & 2.4 min & 1.8 min \\
Forward delay, median                   & 2.3 h & 2.1 h \\
Forward delay, 90th percentile          & 22.0 h & 21.2 h \\
\bottomrule
\end{tabular}
\end{table}

Three properties of Table~\ref{tab:periods} define what the twin must
reproduce. First, the shock is large: on 13 June Gaza/Iran volume reached ten
times its median held-out day, far above anything in the calibration period.
Second, forwarding is fast: half of all forwards occur within about two hours
and one in ten within roughly two minutes, a speed that is itself relevant for
coordination detection. Third, cascades are heavy-tailed and grew during the
crisis: cascades reaching four or more channels were 61\% more frequent in the
held-out period (550 versus 342), even though the number of cascades of size two
or more was nearly unchanged.

\subsection{Known Biases and Limitations of the Corpus}
Four properties of the corpus constrain what can be concluded from it. First,
even after balancing, the frame is shaped by Telegram's forwarding culture:
channels that forward heavily are easier to discover, and these are
predominantly pro-Russian and alternative-geopolitics channels, whereas
mainstream Western outlets have a small Telegram presence. We therefore label
each channel's perspective and report detector errors per perspective rather
than claiming a representative sample of all Telegram discourse on the
conflict. Second, the corpus contains English and Russian text only; Arabic,
Hebrew and Persian channels, which carry much of the conflict's Telegram
discourse, are out of scope. This choice has a measurable effect on
perspective balance. After labelling, Israeli channels carried only 5\% of
Gaza/Iran messages, so we ran an additional targeted search for Israeli news
outlets (i24NEWS, Ynet, Kan, Israel Hayom, Arutz Sheva, Haaretz and others)
under the same filters. No new channel qualified: the large Israeli news
channels on Telegram publish in Hebrew (the largest match, Ynet's verified
Hebrew channel, has over 110{,}000 subscribers), whereas their English-language
Telegram channels have fewer than 1{,}000 subscribers each. The
under-representation of the Israeli perspective is therefore a consequence of
the language scope rather than of the discovery procedure, and extending the
corpus to Hebrew and Arabic is a direct avenue for future work. Third, view, forward and reaction counts were read
at collection time (September 2026), more than a year after publication, and
therefore reflect cumulative rather than contemporaneous exposure. Fourth, a
substantial share of messages carry media without text; these count towards
activity statistics but their content cannot be reproduced by a text-only
simulator.

\section{TeleTwin: Simulator Design}
\label{sec:twin}
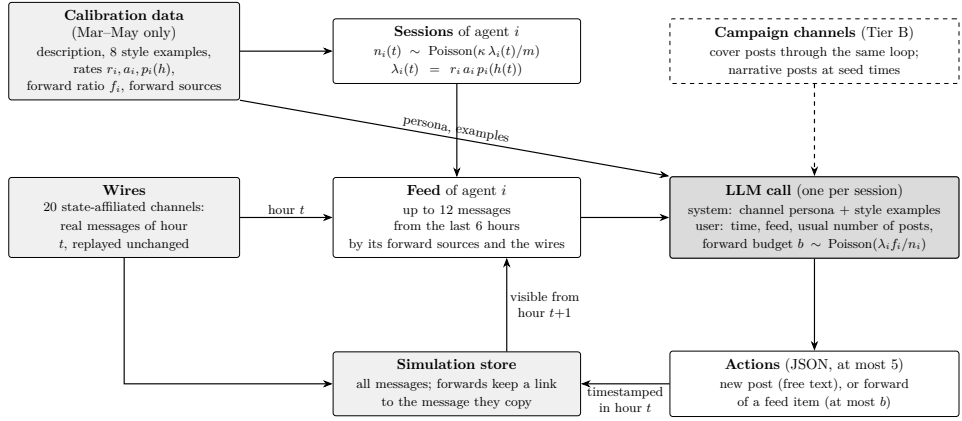
\begin{figure}[t]
\centering
\resizebox{\textwidth}{!}{
\begin{tikzpicture}[
  font=\small,
  box/.style={draw, rounded corners=2pt, align=center, inner sep=5pt},
  data/.style={box, fill=black!6},
  llm/.style={box, fill=black!14, line width=0.8pt},
  dashedbox/.style={box, dashed},
  arr/.style={-{Stealth[length=2.4mm]}, thick},
  lbl/.style={font=\footnotesize, align=center, inner sep=2pt},
]
\node[data, text width=52mm] (calib) at (0,0) {\textbf{Calibration data} (Mar--May only)\\[2pt]
  \footnotesize description, 8 style examples,\\ \footnotesize rates $r_i, a_i, p_i(h)$,\\
  \footnotesize forward ratio $f_i$, forward sources};
\node[data, text width=52mm] (wires) at (0,-4.0) {\textbf{Wires}\\ \footnotesize 20 state-affiliated channels:\\[1pt]
  \footnotesize real messages of hour $t$, replayed unchanged};

\node[box, text width=56mm] (sessions) at (8.0,0) {\textbf{Sessions} of agent $i$\\[2pt]
  \footnotesize $n_i(t) \sim \mathrm{Poisson}(\kappa\,\lambda_i(t)/m)$\\
  \footnotesize $\lambda_i(t) = r_i\, a_i\, p_i(h(t))$};
\node[box, text width=56mm] (feed) at (8.0,-4.0) {\textbf{Feed} of agent $i$\\[2pt]
  \footnotesize up to 12 messages from the last 6 hours\\ \footnotesize by its forward sources and the wires};
\node[data, text width=56mm] (store) at (8.0,-8.0) {\textbf{Simulation store}\\[2pt]
  \footnotesize all messages; forwards keep a link\\ \footnotesize to the message they copy};

\node[dashedbox, text width=66mm] (camp) at (16.6,0) {\textbf{Campaign channels} (Tier B)\\[2pt]
  \footnotesize cover posts through the same loop;\\ \footnotesize narrative posts at seed times};
\node[llm, text width=66mm] (llm) at (16.6,-4.0) {\textbf{LLM call} (one per session)\\[2pt]
  \footnotesize system: channel persona + style examples\\
  \footnotesize user: time, feed, usual number of posts,\\
  \footnotesize forward budget $b \sim \mathrm{Poisson}(\lambda_i f_i / n_i)$};
\node[box, text width=66mm] (actions) at (16.6,-8.0) {\textbf{Actions} (JSON, at most 5)\\[2pt]
  \footnotesize new post (free text), or forward\\ \footnotesize of a feed item (at most $b$)};

\draw[arr] (calib) -- (sessions);
\draw[arr] (calib.south east) -- node[lbl, sloped, above, pos=0.5] {persona, examples} (llm.north west);
\draw[arr] (sessions) -- (feed);
\draw[arr] (wires) -- node[lbl, above] {hour $t$} (feed);
\draw[arr] (wires.south) |- (store.west);
\draw[arr] (feed) -- (llm);
\draw[arr] (llm) -- (actions);
\draw[arr] (actions) -- node[lbl, below] {timestamped\\ in hour $t$} (store);
\draw[arr] ([xshift=12mm]store.north) -- node[lbl, right] {visible from\\ hour $t{+}1$} ([xshift=12mm]feed.south);
\draw[arr, dashed] (camp) -- (llm);
\end{tikzpicture}}
\caption{One simulated hour in TeleTwin. Wires replay real messages; each agent
opens a calibrated number of sessions, and in each session the LLM, conditioned on
the channel's persona and real style examples, reads a feed of recent messages
and returns posts and forwards. The forwarding \emph{rate} is fixed by a
calibrated budget, while the \emph{choice} of what to forward is the model's.
New messages become visible to other agents from the next hour. All rates and
style examples come from the calibration period only. Dashed: campaign channels
added in Tier~B of RQ2.}
\label{fig:loop}
\end{figure}

TeleTwin simulates the Gaza/Iran ecosystem hour by hour. Its design follows
from Telegram's structure: channels, not individual users, are the actors;
content reaches other channels through forwarding; and events enter the
ecosystem largely through a small number of official and state-media sources.
Figure~\ref{fig:loop} summarises one simulated hour.

\subsection{Agents and Wires}
\label{sec:agents}
The twin contains the 124 channels that published at least 100 Gaza/Iran
messages in the collection window. We divide them into two roles.

\paragraph{Wires.} The 20 state-affiliated channels among them (official
bodies and state media; Section~\ref{sec:data}) are not simulated. Their real
messages are replayed unchanged, at their real timestamps, and act as the
exogenous input that carries events into the twin, including the June 2025
war. Wires comprise 14 Russian state outlets and officials, four Iranian state
outlets, the official channel of the Israel Defense Forces, and Al Jazeera
English. Treating them as exogenous mirrors their role on the platform: they
originate reports that the rest of the ecosystem reacts to, and they rarely
forward other channels. The wire set was fixed when the twin was built; one
channel later relabelled as state-affiliated during the authors' review of the
perspective labels (SouthFront) therefore remains a simulated agent, while all
per-perspective analyses use the reviewed labels.

\paragraph{Agents.} The remaining channels are simulated by an LLM. Four of
them published nothing before June and cannot be instantiated from the
calibration period, which leaves 100 agents: 51 pro-Russian, 15
alternative-geopolitics, 12 Palestinian/Arab, 6 Iranian-axis, 6 OSINT
aggregators, 6 Israeli, 2 pro-Ukrainian and 2 US right-wing channels; 57 write
mainly in English and 43 in Russian. Each agent is instantiated exclusively from
its own calibration-period (March--May) data:
\begin{itemize}[leftmargin=*]
  \item its title and public description;
  \item eight of its own posts as style examples, with the share of Gaza/Iran
    examples matching the share of Gaza/Iran messages in its own calibration
    posts (Section~\ref{sec:tuning} explains why this matters);
  \item its dominant language;
  \item its activity rate: messages per active day, the share of days on which
    it was active, and its hour-of-day profile;
  \item its forward ratio and the up to 15 channels it forwarded from most, with
    their relative frequencies.
\end{itemize}
The agent's perspective label is \emph{not} given to the model: stance enters
only through the channel's own description and posts, so that the twin
reproduces how a channel actually writes rather than a stereotype of its
category.

\subsection{One Simulated Hour}
\label{sec:loop}
At each hour $t$ the twin performs three steps.

\paragraph{1. Wire replay.} All real wire messages with timestamps in $t$ are
published.

\paragraph{2. Agent sessions.} For agent $i$, the expected number of messages in
hour $t$ is
\begin{equation}
  \lambda_i(t) = r_i \, a_i \, p_i\big(h(t)\big),
\end{equation}
where $r_i$ is its calibration messages per active day, $a_i$ the share of
calibration days on which it was active, and $p_i(h)$ the share of its messages
published at hour of day $h$. Agent $i$ opens
$n_i(t) \sim \mathrm{Poisson}\big(\kappa\,\lambda_i(t)/m\big)$ sessions, where
$m = 1.9$ is the mean number of messages an LLM call produces and $\kappa=1.25$
is a scale factor tuned on calibration data (Section~\ref{sec:tuning}).
Busy channels therefore open several sessions per hour, and rarely active
channels usually open none. Each session is one LLM call.

In a session the model receives a system prompt with the channel's title,
description, language and style examples, and an instruction to stay fully in
character. The user prompt gives the current date and time, a \emph{feed} of
up to 12 messages from the previous six hours, published by the channels the
agent forwarded from during calibration or by wires, and the number of own
posts the channel usually publishes in such a session. The prompt states that
the channel may publish more or less than usual, or nothing, if the situation
calls for it. The model replies in JSON with up to five actions, each either a
new post (free text) or a forward of a feed item.

\paragraph{Forward budget.} In tuning, the forwarding rate proved highly
sensitive to prompt wording: small changes moved it between 2\% and 24\% of
messages against a real rate of about 10\%. We therefore separate the
\emph{rate} of forwarding, which is calibrated, from the \emph{choice} of what to
forward, which is left to the model. Each session receives a forward budget
$b \sim \mathrm{Poisson}\big(\lambda_i(t)\,f_i / n_i(t)\big)$, where $f_i$ is
the agent's calibration forward ratio; the prompt asks for exactly $b$ forwards
(or none), and forwards beyond the budget are discarded. The model thus decides
which items are worth forwarding, which is what shapes cascades, while the
overall forwarding rate stays anchored to the real channel.

\paragraph{3. Publication.} Each action is timestamped at a random minute
within $t$ and written to the simulation store. Forwards keep a link to the
message they copy, so cascades can be reconstructed exactly as for the real
corpus. New messages become visible in other agents' feeds from hour $t+1$.

Crucially, nothing in the prompt tells an agent that a crisis is under way or
that it should post more. Volume, topics and forwarding respond to the shock
only if the agents, reading the wire messages and each other's posts, choose to
respond. Whether the June spike emerges is therefore an outcome of the
simulation, not an input.

\paragraph{What is not simulated.} TeleTwin does not model subscribers, views or
reactions, and it does not represent forwards from channels outside the frame,
which carry about half of the real forwards (Section~\ref{sec:data}). It also
generates text only, so the 22\% of real messages that carry media without text
have no counterpart.

\subsection{Model and Implementation}
All agents are played by Qwen2.5-72B-Instruct (AWQ 4-bit quantisation), served
with vLLM on a single NVIDIA H200 GPU, with temperature 0.9 and at most 600
output tokens per call. We use an open-weights model so that the twin can be
run locally and reproduced without commercial APIs. Simulating one hour takes
between about 20 seconds at night and 60 seconds at peak hours, so the
three-month held-out period requires roughly 29 GPU-hours per run.

\subsection{Calibration Protocol}
\label{sec:tuning}
All tuning used calibration-period windows only; the held-out period was not
consulted. We tuned on three windows in May 2025 (20 May 12:00--18:00, 8 May
06:00--12:00 and 26 May 00:00--12:00), comparing simulated and real volume and
forward share per hour. Because the busiest real day can differ substantially
from an average one (8 May was 23\% above the calibration average for the same
hours), the target was the average-day expectation implied by the agents'
calibration rates rather than the volume of any single day. Seven runs were
needed to calibrate volume and forwarding, and two further runs addressed topic
over-focus (Appendix~\ref{app:tuning}). Two findings shaped the final design: the
forwarding rate cannot be set reliably through prompt wording, which led to the
forward budget; and a single LLM call produces a roughly fixed amount of
output, which led to multiple sessions per hour for busy channels. On the final
check, a twelve-hour window not used in earlier runs, the twin produced
1.01 times the average-day expectation (0.93 times the real volume of that
day), with 7\% forwards against 9\% in the real data, and followed the real
night-time low and morning rise.

\paragraph{Topic over-focus.} The first six tuning runs targeted volume and
forwarding but not topic. The resulting configuration (which we call v1) chose
each agent's style examples to be half about Gaza/Iran where possible, to
ensure that every agent had on-topic examples. A first analysis of the held-out
run revealed that agents wrote about Gaza/Iran far more often than their real
counterparts, and we confirmed the problem on the 26 May calibration window:
52\% of simulated messages concerned Gaza/Iran, against 19\% of real ones.
The feed was not the cause, since only 19\% of wire messages in calm periods
concern Gaza/Iran; the over-focus arises on the agent side. We tested two
corrections on the same calibration window, keeping all other settings fixed.
Making each agent's style examples proportional to its real topic mix (v2)
reduced the Gaza/Iran share to 40\% without affecting volume (1.01 times the
average-day expectation) or forwarding (8\% against 9\%). Additionally stating
the channel's usual Gaza/Iran share in the prompt (v3) had no further effect
(42\%) and raised volume to 1.19 times the expectation, and was discarded. As
with forwarding, the model does not follow a rate stated in the prompt. We use
v2 as the main configuration and report v1 for comparison. Because the topic
problem was discovered on held-out data, we state this explicitly: the v1 to v2
change was diagnosed with the held-out run but tested and adopted on
calibration data only, and it changes no parameter fitted to the held-out
period. The remaining over-focus of v2 (about twice the real share) is a
limitation of the twin that we quantify in Section~\ref{sec:rq1}.

\subsection{Baselines}
\label{sec:baselines}
To establish what the LLM adds, we compare TeleTwin with two non-LLM null
models that use the same agents, calibration statistics and output format.
\begin{itemize}[leftmargin=*]
  \item \textbf{Static.} Each agent publishes
    $\mathrm{Poisson}(\lambda_i(t))$ messages per hour. Each message is a
    forward with probability $f_i$, drawn from the recent messages of its
    calibration forward sources, and is otherwise a post whose text is
    resampled from the agent's own calibration-period posts.
  \item \textbf{Reactive.} As the static model, but the hourly rate is
    multiplied by $\big(w(t)/\bar{w}\big)^{\beta}$, where $w(t)$ is the number
    of wire messages in hour $t$ and $\bar{w}$ its calibration mean. The
    elasticity $\beta$ is fitted on calibration days by least squares on log
    daily volumes, giving $\beta = 0.50$. This baseline encodes the simple
    hypothesis that the ecosystem posts more when official sources post more;
    the LLM twin must beat it to show that its response to events is more than
    a volume multiplier.
\end{itemize}

\section{RQ1: Fidelity}
\label{sec:rq1}

\subsection{Protocol}
We run TeleTwin and both baselines over the held-out period (1 June -- 31
August 2025) and compare each with the real messages of the same 100 agent
channels. We report two TeleTwin configurations: v1 (half Gaza/Iran style
examples) over the full held-out period, and v2 (proportional style examples)
over 1--24 June, which covers one pre-war week and the full twelve-day war. Wires are identical in all runs, because they are replayed from real
data. Simulated messages are tagged with the same Gaza/Iran keyword tagger as
the real corpus (Section~\ref{sec:tagging}).

\subsection{Metrics}
We evaluate fidelity along five dimensions.
\begin{enumerate}[leftmargin=*]
  \item \textbf{Shock response.} Daily Gaza/Iran volume over the held-out
    period: its correlation with the real series, and the peak ratio on 13
    June relative to the median of the preceding week. In the real data this
    ratio is about ten.
  \item \textbf{Topics.} Jensen--Shannon divergence between the topic
    distributions of real and simulated messages, using a topic model fitted
    on the real calibration-period corpus and applied to both.
  \item \textbf{Cascades.} Kolmogorov--Smirnov distance between the real and
    simulated distributions of cascade size (the number of agents that forward
    the same original message), and the change in large cascades between the
    calibration and held-out periods, which grew by 61\% in the real data.
  \item \textbf{Forwarding network.} Overlap (Jaccard) between real and
    simulated forward edges among agents and wires, and their degree
    distributions.
  \item \textbf{Indistinguishability.} The area under the ROC curve of a
    classifier trained to separate real from simulated posts of the same
    channel and period; an AUC close to 0.5 indicates that simulated posts
    cannot be told apart from real ones. We report it per language and per
    perspective.
\end{enumerate}
For every metric we report TeleTwin alongside the static and reactive
baselines, and we describe failure modes, for example events invented by
agents that do not appear in any wire message, as well as matches.

\subsection{Results}
Table~\ref{tab:rq1} compares TeleTwin (configuration v1) and the two baselines
with the real messages of the 100 agent channels over the whole held-out period.

\begin{table}[t]
\centering
\small
\caption{RQ1: fidelity over the held-out period (1 June -- 31 August 2025),
100 agent channels. Arrows indicate the better direction. The classifier AUC
is the area under the ROC curve for separating real from simulated posts of the
same channels (0.5 = indistinguishable). Baselines: mean of three runs
(standard deviations are below 0.01 except where shown); TeleTwin v1: one run.}
\label{tab:rq1}
\setlength{\tabcolsep}{4pt}
\begin{tabular}{lrrrr}
\toprule
 & TeleTwin (v1) & Reactive & Static & Real \\
\midrule
Volume (simulated / real)                     & 0.91 & 0.94 & 0.88 & 1 \\
Forward share                                 & 6.9\% & 10.3\% & 10.1\% & 8.9\% \\
Gaza/Iran share                               & 52.1\% & 22.1\% & 22.5\% & 26.7\% \\
Daily Gaza/Iran correlation $\uparrow$        & \textbf{0.86} & 0.64 {\scriptsize$\pm$0.05} & $-0.08$ {\scriptsize$\pm$0.09} & 1 \\
Peak ratio, 13 June $\uparrow$                & \textbf{2.19} & 1.40 {\scriptsize$\pm$0.04} & 0.98 {\scriptsize$\pm$0.05} & 10.62 \\
Topic JSD $\downarrow$                        & 0.084 & 0.018 & \textbf{0.016} & 0 \\
Cascade-size KS $\downarrow$                  & 0.080 & 0.073 & \textbf{0.072} & 0 \\
Cascades of size $\geq 4$                     & 3.2\% & 0.1\% & 0.1\% & 1.1\% \\
Largest cascade (channels)                    & 12 & 5 & 5 & 6 \\
Forward edges                                 & 1{,}484 & 458 & 457 & 679 \\
Forward-edge Jaccard $\uparrow$               & 0.22 & \textbf{0.43} & \textbf{0.43} & 1 \\
In-degree rank correlation $\uparrow$         & 0.84 & \textbf{0.92} & 0.91 & 1 \\
Classifier AUC $\downarrow$                   & 0.96 & 0.65 & 0.66 & 0.5 \\
\bottomrule
\end{tabular}
\end{table}

\begin{figure}[t]
\centering
\includegraphics[width=\textwidth]{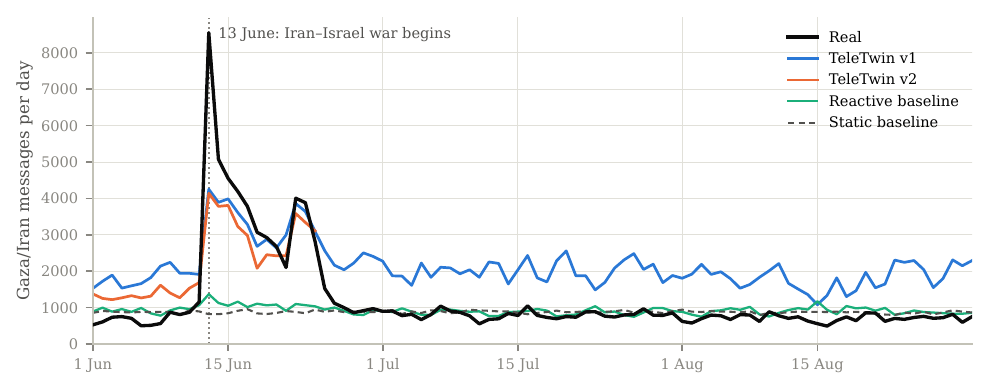}
\caption{Daily Gaza/Iran messages of the 100 agent channels, 1 June -- 31 August
2025: real data, TeleTwin (v1 over the whole period, v2 over 1--24 June) and the
two baselines. The dotted line marks 13 June, the first day of the Iran--Israel
war. No simulated agent is told that a war has started.}
\label{fig:daily}
\end{figure}

\paragraph{Response to events.} TeleTwin is the only model whose dynamics
follow the conflict (Figure~\ref{fig:daily}). Its daily Gaza/Iran volume correlates with the real series
at $r = 0.86$, against $0.64 \pm 0.05$ (three runs) for the reactive baseline, which scales activity
with wire volume, and $-0.16$ for the static baseline. On 13 June, the first
day of the Iran--Israel war, the twin's Gaza/Iran volume more than doubled
(2.2 times its median of the preceding week) without any instruction to react,
and in the following week it tracked the real decline closely, within roughly
2--25\% of the real daily volume. The response is nevertheless far too small on
the first day: the real ecosystem's volume rose 10.6-fold, and the twin reached
about half of the real absolute volume on that day.

\paragraph{Volume and forwarding.} Aggregate volume is close to real (0.91),
and the forward share (6.9\%) is somewhat below the real 8.9\%; both follow from
calibration. The twin reproduces which channels are central: the rank
correlation of agents' in-degree in the forwarding network is 0.84 and of their
out-degree 0.85. It is less faithful about \emph{whom} channels forward from.
Only 22\% of forward edges coincide with real ones, against 43\% for the
baselines, which sample forward sources from each channel's calibration
habits, and the twin creates more than twice as many distinct forward edges as
the real ecosystem. Agents choose items from their feed by content rather than
by habitual source.

\begin{figure}[t]
\centering
\includegraphics[width=0.62\textwidth]{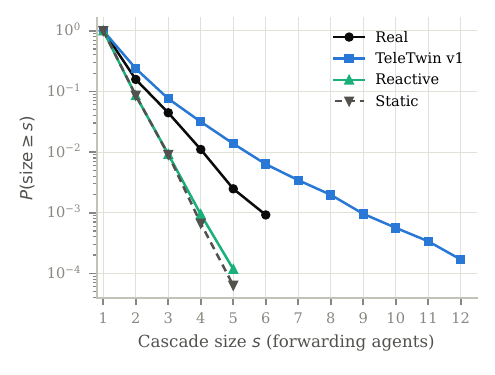}
\caption{Distribution of cascade sizes over the held-out period (share of
forwarded messages whose cascade reaches at least $s$ agents, log scale). The
twin's cascades are too large; the baselines' are too small.}
\label{fig:cascades}
\end{figure}

\paragraph{Over-amplification.} Cascades in the twin are too large
(Figure~\ref{fig:cascades}). Messages
forwarded by four or more agents make up 3.2\% of cascades, against 1.1\% in the
real data, and the largest cascade reaches 12 channels, against 6. The baselines
err in the opposite direction (below 0.1\%), since their forwarding choices are
independent. The two errors bracket the real value; the twin's agents converge
on the same salient items more than real channels do.

\paragraph{Topics and surface realism.} Half of the twin's messages concern
Gaza/Iran, against a quarter of the real messages: the topic over-focus
identified in calibration (Section~\ref{sec:tuning}) persists over the whole
period, and it also explains part of the small peak ratio, since a high
baseline leaves less room for a spike. The topic distribution of the twin is
further from the real one (JSD 0.084) than that of the baselines (0.016--0.018),
which is expected because the baselines resample real calibration-period posts.
Finally, a character n-gram classifier separates the twin's posts from real
posts of the same channels with AUC 0.96, consistently across English (0.96)
and Russian (0.96) and across perspectives (0.92--0.98). The reference point is
not 0.5: posts resampled from the calibration period, which are real but cannot
follow later events, are separated with AUC 0.66. LLM-generated posts thus
remain recognisable as such, which is relevant for RQ2: in real data,
LLM-written campaigns may be detectable by style as well as by coordination.

\paragraph{Effect of proportional style examples.} Table~\ref{tab:v2}
compares configurations v1 and v2 over 1--24 June. Making style examples
proportional to each channel's real topic mix lowered the Gaza/Iran share from
64\% to 55\% (real: 38\%) and raised the peak ratio on the first day of the war
from 2.2 to 3.0, while leaving volume, forwarding, the daily correlation and
all structural metrics essentially unchanged. This supports the interpretation
that the high baseline attention to the conflict is one of the causes of the
weak shock response: the less the agents talk about Gaza/Iran on ordinary days,
the larger their reaction to the war. The remaining gap (3.0 against 10.6) shows
that over-focus is not the only cause.

\begin{table}[t]
\centering
\small
\caption{Configurations v1 and v2 over 1--24 June 2025 (pre-war week and war),
with the baselines for reference. v2 differs from v1 only in the topic mix of
the agents' style examples. Baselines: mean of three runs; v1 and v2: one
run each.}
\label{tab:v2}
\begin{tabular}{lrrrrr}
\toprule
 & v1 & v2 & Reactive & Static & Real \\
\midrule
Volume (simulated / real)              & 0.67 & 0.66 & 0.74 & 0.64 & 1 \\
Gaza/Iran share                        & 63.6\% & 55.0\% & 22.3\% & 22.4\% & 38.1\% \\
Daily Gaza/Iran correlation $\uparrow$ & 0.93 & \textbf{0.93} & 0.81 & $-0.18$ & 1 \\
Peak ratio, 13 June $\uparrow$         & 2.19 & \textbf{2.96} & 1.40 & 0.98 & 10.62 \\
Topic JSD $\downarrow$                 & 0.085 & 0.093 & 0.054 & \textbf{0.052} & 0 \\
Cascades of size $\geq 4$              & 3.8\% & 4.3\% & 0.09\% & 0.10\% & 1.1\% \\
Forward-edge Jaccard $\uparrow$        & 0.22 & 0.20 & \textbf{0.43} & \textbf{0.43} & 1 \\
Classifier AUC $\downarrow$            & 0.97 & 0.96 & 0.75 & 0.74 & 0.5 \\
\bottomrule
\end{tabular}
\end{table}

\paragraph{Summary.} TeleTwin captures what the statistical baselines cannot:
the timing of the ecosystem's response to events and the identity of its hubs.
It is weaker than the baselines on surface realism (topic mix, text style) and
on the micro-structure of forwarding (sources and cascade sizes). We consider
these failure modes as findings in their own right; they are discussed in
Section~\ref{sec:discussion}.

\section{RQ2: Detecting LLM-Powered Campaigns}
\label{sec:rq2}
RQ2 asks how well coordination detectors find LLM-powered campaigns whose
membership is known exactly. We construct campaigns, inject them into Telegram
data, and measure detection at the level of channels. We use two settings.
In \textbf{Tier A}, campaigns are injected into the \emph{real} held-out data;
the organic background is real, and only the campaign is synthetic. In
\textbf{Tier B}, campaign channels are agents inside TeleTwin, so organic
agents can encounter, forward and restate campaign content.

\subsection{Campaign Construction}
A campaign consists of $K$ synthetic channels. To give each one a credible
cover, it borrows the voice of a real agent channel (its title, description and
style examples; Section~\ref{sec:agents}), and most of what it publishes is
\emph{filler}: ten ordinary posts per day in that voice, written by the LLM in
reaction to the real wire headlines of the same day. On top of this cover, each
channel publishes \emph{narrative} posts that push the campaign's message at
four seed times per day.

\paragraph{Narratives.} All narratives are fictional. One set (\emph{named})
concerns invented organisations and outlets, for example a fictitious relief
group said to divert aid convoys; a second set (\emph{generic}) consists of
vague, mild claims with no distinctive name, for example rumours that a new
humanitarian corridor will open. Using fictional content means that the
benchmark never generates disinformation about real people or groups, and it
does not weaken the test: coordination detectors measure \emph{how} channels
coordinate, not whether a claim is true. Generic narratives remove the shared
rare names that make named narratives easy to match.

\paragraph{Design factors.} Campaigns vary along five factors:
\begin{itemize}[leftmargin=*]
  \item \textbf{Style.} \emph{Copy}: every channel posts the same seed text.
    \emph{Paraphrase}: every channel posts its own LLM rewrite of the seed.
    \emph{Cross-lingual}: as paraphrase, with half of the channels writing in
    Russian and half in English. \emph{Persona}: each channel restates the
    claim in its own borrowed voice.
  \item \textbf{Timing.} \emph{Synchronised}: all channels post within five
    minutes of the seed time. \emph{Staggered}: posting times are spread over
    six hours.
  \item \textbf{Scale.} $K \in \{5, 10, 20\}$ channels.
  \item \textbf{Links.} Posts either carry the narrative's URL (a reserved
    \texttt{.example} domain) or not.
  \item \textbf{Context and intensity.} A calm window (1--14 July 2025) or a
    crisis window (13--26 June 2025, the first two weeks of the war, when
    organic volume is highest); and four or one narrative posts per channel per
    day.
\end{itemize}
We run 20 configurations: the full style $\times$ timing grid at $K=10$ in the
calm window, a scale sweep, link variants, three crisis variants, and five
harder variants with generic narratives, including two with one narrative post
per day (the full list is in the released configuration). Campaign text is
generated by the same model that plays the TeleTwin agents, so that in Tier B
detectors cannot separate campaigns from organic agents by the model's style
alone.

\subsection{Detectors}
\label{sec:detectors}
We evaluate five detector families that cover the main signals used in
coordination detection, all operating on public message streams. Each assigns
a score to every channel.
\begin{itemize}[leftmargin=*]
  \item \textbf{Identical text} (\emph{co\_text}): two channels post the same
    normalised text within one hour. Telegram forwards are excluded, since
    they carry visible attribution.
  \item \textbf{Co-URL} (\emph{co\_url}): two channels post the same URL within
    one hour, the coordinated-link-sharing signal of Giglietto et
    al.~\cite{giglietto2020takes}.
  \item \textbf{Synchrony} (\emph{sync}): two channels post in the same
    two-minute bins more often than their activity levels predict.
  \item \textbf{Lexical similarity} (\emph{lexical}): near-duplicate posts,
    defined as word uni- and bigram TF-IDF cosine similarity of at least 0.6
    within six hours.
  \item \textbf{Semantic similarity} (\emph{semantic}): posts whose
    multilingual sentence embeddings~\cite{reimers2019sentencebert}
    (paraphrase-multilingual-MiniLM-L12-v2) have cosine similarity of at least
    0.85 within six hours. This model maps English and Russian into a shared
    space, so it can match a post with its translation.
\end{itemize}
The pairwise detectors share one aggregation step, following the
network-based approach of Pacheco et al.~\cite{pacheco2021uncovering}. For each
pair of channels we count coordinated message pairs $c_{ij}$ and normalise by
volume, $w_{ij} = c_{ij}/\sqrt{n_i n_j}$, where $n_i$ is the number of original
posts of channel $i$. The score of channel $i$ is the sum of its five strongest
ties. Normalisation is essential in a news ecosystem: in our first tests,
unnormalised counts ranked large organic channels above campaign channels,
because channels covering the same events inevitably publish many similar
posts. A campaign, by contrast, is a small group that shares an unusually large
\emph{fraction} of its output.

\subsection{Evaluation}
In Tier A, the detection population is every channel of the corpus active in
the campaign window (up to 195 organic channels) plus the $K$ campaign
channels. We report ROC-AUC and average precision for separating campaign from
organic channels, and precision among the $K$ highest-scoring channels.
Because a detector that flags one side's ordinary channels more than another's
would be learning politics rather than coordination, we also fix, for each
detector, the threshold that recovers 80\% of the campaign channels and report
the share of organic channels of each perspective that it wrongly flags.

\paragraph{Tier B setup.} In Tier B, the $K$ channels of a campaign are added
to TeleTwin as agents. Their cover posts are not pre-generated: each campaign
channel is simulated like any other agent, with the borrowed voice of its donor
channel and the donor's calibrated activity rate, so its cover traffic responds
to the same feed as the organic agents. Its narrative posts are the same texts
as in Tier A, injected at their scheduled times. In the \emph{isolated}
condition no organic agent follows the campaign channels, so their posts never
appear in organic feeds. In the \emph{embedded} condition each campaign channel
is added to the forward sources of five organic agents, preferably of the
donor's perspective, which mimics a campaign that has built a small audience
among like-minded channels. Campaign channels appear in feeds under
ordinary-looking names. We run the isolated and embedded conditions for a
paraphrase campaign, the embedded condition for the harder persona campaign
with generic narratives, and a control run without any campaign, all over the
calm window with configuration v2. Besides detection, Tier B lets us measure
\emph{diffusion}: how often organic agents forward campaign posts, and how often
they publish original posts that restate a campaign narrative (semantic
similarity of at least 0.80 to a narrative post of the previous 24 hours),
which we call narrative laundering. We also check whether organic channels
that restated a narrative appear among the detectors' false positives.

\subsection{Results: Tier A}
Figure~\ref{fig:tierA} summarises ROC-AUC for every campaign and detector;
Table~\ref{tab:tierA} in Appendix~\ref{app:tierA} also gives average precision
(AP) and precision among the top $K$ channels for the semantic detector. Each campaign was generated three times, with a
different set of cover channels and independently generated text, and we report
means over the three versions; results are highly consistent across versions.

\begin{figure}[t]
\centering
\includegraphics[width=0.75\textwidth]{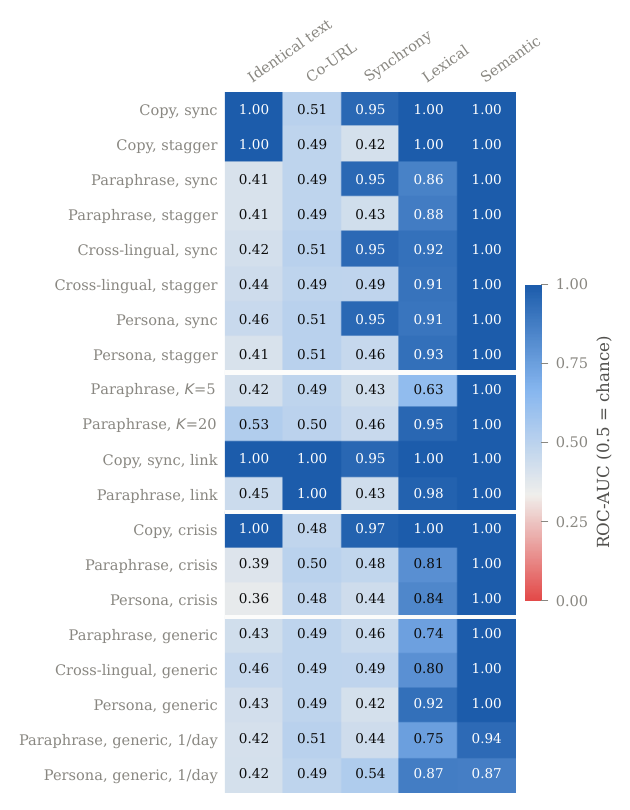}
\caption{RQ2 Tier A: ROC-AUC of each detector (columns) for each injected
campaign (rows), among up to 195 real channels; mean over replicate campaigns
with different cover channels and newly generated text. Blue: detected; gray:
chance (0.5); red: below chance. Rows are grouped into the main style $\times$
timing grid, scale and link variants, crisis-window variants and harder
variants. Numerical values, average precision and precision@K are given in
Table~\ref{tab:tierA} (Appendix~\ref{app:tierA}).}
\label{fig:tierA}
\end{figure}

\paragraph{Classic signals fail against simple LLM strategies.} Each classic
detector works only against the strategy it was designed for. Identical-text
detection identifies copy-paste campaigns perfectly and falls to chance (AUC
0.36--0.53) as soon as each channel posts its own paraphrase. Synchrony
detection reaches AUC 0.95--0.97 when campaign channels post within five
minutes of each other, and falls to 0.42--0.54 when posting is spread over six
hours. Co-URL detection is perfect when campaign posts carry a shared link and
uninformative otherwise. All three signals can therefore be removed at no cost
by an operator using an LLM: rewording, staggering and dropping the link require
no additional effort.

\paragraph{Lexical similarity degrades.} Near-duplicate detection on word
n-grams survives paraphrase partly (AUC 0.86--0.93 for named narratives) but
degrades with every factor that reduces lexical overlap: small campaigns (0.63
at $K=5$), crisis-time background volume (0.81), and generic narratives without
distinctive names (0.74--0.80).

\paragraph{Semantic similarity works, until campaigns slow down.}
Multilingual semantic similarity identifies every campaign that publishes four
narrative posts per channel per day, with AUC 1.00 and precision@K of 0.87 or
more, regardless of style, timing, language mixing, scale, crisis volume or
narrative type. Cross-lingual campaigns, in which half of the channels write in
Russian, are found as easily as monolingual ones. The picture changes for
low-intensity campaigns. With one narrative post per channel per day and
generic narratives, AUC remains high (0.94 for paraphrase, 0.87 for persona)
but average precision drops to 0.32 and 0.18, and precision among the top ten
channels to 0.10 and 0.03: campaign channels are ranked above most organic
channels but not above the organic channels that are themselves most
similar to each other. In practical terms, an analyst reviewing the ten most
suspicious channels would find at most one campaign channel.

\paragraph{False positives are not evenly distributed.} At the threshold that
recovers 80\% of campaign channels, the semantic detector flags at most 0.3\% of
organic channels for any campaign with four narrative posts per day. For the
low-intensity persona campaign, it flags 16.6\% of organic channels on average,
and these errors fall unevenly: 30\% of the alternative-geopolitics channels,
29\% of OSINT aggregators and 27\% of Iranian-axis channels, against 16\% of
pro-Russian and 2\% of Palestinian/Arab channels and no Israeli channel. The
lexical detector shows the same unevenness more strongly: in the crisis window
it flags 62.5\% of Israeli channels and 56\% of OSINT aggregators, against
8\% of Palestinian/Arab channels. Some perspective groups are small (eight
Israeli and two US right-wing channels), so these rates are indicative rather
than precise. They nevertheless show that the organic channels most likely to be
mistaken for a campaign are those whose own output is internally homogeneous,
such as aggregators that relay the same agencies and state-aligned outlets that
echo each other. A detector tuned to catch quiet campaigns would therefore
disproportionately flag particular political communities.

\subsection{Results: Tier B}
We evaluate the three Tier B conditions over their first week (1--7 July),
which contains 28 narrative posts per campaign channel, with two independent
runs per condition (different random seeds and model samples). We report both
runs where they differ.

\paragraph{All detectors fail inside the twin.} Inside TeleTwin, no detector
separates campaign channels from organic agents. For the isolated paraphrase
campaign, which every semantic test in Tier A identified perfectly, the
semantic detector reaches ROC-AUC 0.61 and 0.58 in the two runs (AP 0.10 and
0.09), and the lexical detector 0.63 and 0.60; the identical-text, co-URL and
synchrony detectors are at or below chance. Results are similar for the embedded
paraphrase campaign (semantic AUC 0.61 and 0.59) and the embedded persona
campaign with generic narratives (0.57 and 0.58). At the threshold that recovers
80\% of the campaign channels, the semantic detector flags 62--72 of the 120
organic channels across the six runs.

\begin{figure}[t]
\centering
\includegraphics[width=0.62\textwidth]{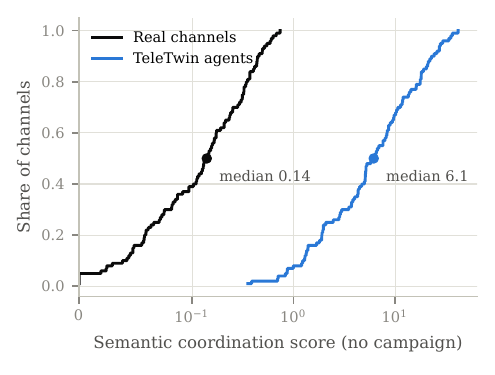}
\caption{Semantic coordination scores of organic channels, with no campaign
present, 1--7 July 2025: real channels against the corresponding TeleTwin agents
(cumulative distribution, symmetric-log scale). Dots mark the medians. The
simulated background is about 40 times more homogeneous than the real one.}
\label{fig:homogeneity}
\end{figure}

\paragraph{Why: a homogeneous background.} Two differences from Tier A
contribute. The first is dilution: in Tier B, campaign channels inherit their
donor channel's activity rate and publish on average 45.5 cover posts per day,
against 10 in Tier A, alongside the same four narrative posts. The second, and
much larger, is the background itself. Computed on organic channels only, with
no campaign involved, the median semantic coordination score of an agent
channel over the same week is 0.14 in the real data and 6.1 in the twin,
about 40 times higher (90th percentile: 0.44 against 22.9;
Figure~\ref{fig:homogeneity}). The twin's agents
all react to the same wire messages in the idiom of the same model, so they
produce many semantically near-identical posts across channels. To a semantic
coordination detector, the simulated organic ecosystem already looks like a
coordinated campaign, and a real campaign cannot stand out against it.

\paragraph{Narrative laundering.} Embedding campaign channels in organic feeds
caused organic agents to adopt the campaign's narratives. Organic agents
forwarded campaign posts only two to five times (paraphrase) or never
(persona), but they published original posts that restated a campaign narrative
within a day of seeing it: 46 and 31 such posts in the two paraphrase runs, by
15 organic channels in each, and 50 and 24 in the persona runs, by 10 and 12
channels. Most of these channels were followers of a campaign channel (12--13
of 15 for paraphrase, 7--12 for persona). In the isolated condition no organic
agent restated a narrative in either run. Organic channels that restated a
narrative were then among the detectors' false positives: 11 of the 64--66
organic channels flagged by the semantic detector in each embedded paraphrase
run, and 6--9 of 65--72 in the persona runs, were narrative launderers. In other words, the organic channels that a
campaign succeeds in influencing are among those most likely to be mistaken for
its members.

\section{Discussion}
\label{sec:discussion}

\paragraph{What the twin can and cannot do.} TeleTwin reproduces the dynamics
that statistical models cannot: it reacts to a war it was never told about,
tracks the daily rise and fall of attention to the conflict, and keeps the same
channels at the centre of the forwarding network. It does so with one
open-weights model, agents instantiated only from calibration data, and no
parameter fitted to the held-out period. At the same time, its failures are
systematic rather than random, and they share a common root: LLM agents are
more alike than real channels. They over-focus on the most salient conflict,
converge on the same items when they forward, ignore their habitual sources,
and write in a recognisable common idiom. These are the same properties that
make the background homogeneous in Tier B. We expect them to affect other
LLM-based social simulators as well, and we suggest that homogeneity
diagnostics (topic share against calibration, cascade sizes, organic semantic
similarity, real-versus-simulated classifiers) should be reported routinely.

\paragraph{Prompts do not set rates.} Two of our calibration findings concern
how to control LLM agents. Stated rates were ignored: asking for a given share
of forwards or of on-topic posts moved behaviour unpredictably or not at all.
What worked was to separate rates from choices: the forwarding rate, posting
rate and topic mix of the style examples are taken from data, while the model
decides what to forward and what to write. This division of labour between
calibrated statistics and generated content may be a useful design principle
for digital twins more generally.

\paragraph{Detecting LLM-powered campaigns in real data.} Tier A shows that the
classic coordination signals (identical text, shared links, synchrony) offer no
protection against an operator who uses an LLM to reword posts, staggers them
and drops links, all of which cost nothing. Semantic similarity over
multilingual embeddings recovers loud campaigns, including those that mix
languages, so cross-lingual coordination is not by itself an obstacle for
detection. Quiet campaigns are another matter: at one narrative post per
channel per day, campaign channels no longer stand out from the organic
channels that are most similar to each other, and lowering the threshold to
catch them flags whole groups of organic channels, unevenly across political
perspectives. In our data these are mostly aggregators and state-aligned
outlets that relay the same sources. Any operational use of semantic detectors
should therefore report and audit errors by community.

\paragraph{When the background is written by LLMs.} Tier B was designed as a
harder version of Tier A. Instead, it revealed that an LLM-generated
background is so homogeneous that semantic detection collapses. This has two
implications. For research, it means that detection benchmarks built inside
LLM simulations can be misleading unless the homogeneity of the simulated
background is measured and calibrated against real data; our measurement makes
this concrete (a 40-fold difference). For practice, it is a warning about the
future: as more organic content on platforms is written or rewritten with
LLMs, the semantic diversity that today lets campaigns stand out may shrink.
Tier B also shows that campaigns that succeed in being picked up by organic
channels make detection worse in a second way, because the influenced organic
channels resemble the campaign. Distinguishing a campaign from the channels it
has persuaded will require signals beyond content similarity, such as account
history, network position over time and the direction of influence.
\section{Limitations}
\label{sec:limitations}
We group the limitations of this study into those of the data, of the twin and
of the benchmark.

\paragraph{Data.} The corpus covers public Telegram channels only, in English
and Russian, over six months, and its sampling frame is shaped by forwarding:
channels that forward heavily are easier to discover, and these are
predominantly pro-Russian and alternative-geopolitics channels. Balancing
raised the Palestinian/Arab and Iranian share of Gaza/Iran messages to levels
comparable with the pro-Russian share, but Israeli channels carry only 5\% of
them, because large Israeli outlets publish on Telegram in Hebrew
(Section~\ref{sec:data}). Perspective labels were assigned by the authors from
channel descriptions and sample posts; they are coarse and may be contested for
individual channels. View, forward and reaction counts were read a year after
publication.

\paragraph{The twin.} TeleTwin simulates 100 channels, not the platform. It
does not model subscribers, views or reactions, it treats forwards from channels
outside the frame as absent, and it generates text only. Agents have no memory
beyond their style examples and their six-hour feed, so they cannot pursue a
line of argument over days. The twin relies on one open-weights model; other
models may behave differently, and the over-focus and over-amplification we
observe may be partly model-specific. Wires are replayed from real data, so the
twin is a model of how the ecosystem \emph{responds} to official sources, not of
how events arise. The number of replicates differs by experiment: the
baselines were run three times, every Tier A campaign was generated three times
and each Tier B condition was run twice, with consistent results, but the
TeleTwin runs for RQ1 (v1 over three months, v2 over 24 days) are single runs,
because each requires tens of GPU-hours. The Tier B runs cover one week, and
Tier B confounds two factors (cover volume and background homogeneity),
although we measured the second separately on organic channels alone.
Finally, several design choices (the forward budget, sessions per hour,
proportional style examples) anchor rates to calibration data. They make the
twin more faithful but also mean that some quantities, such as the forward
share, are fitted rather than emergent; we have tried to separate clearly
which quantities are fitted and which are outcomes.

\paragraph{The benchmark.} Campaigns are synthetic and fictional. They follow
patterns described in the literature (copied, paraphrased, translated and
persona-based messaging; synchronised and staggered timing; shared links), but
real operations may combine these strategies, adapt to detection, build up
their channels' credibility over months, or use human operators. Campaign
channels in Tier A are new channels with two weeks of history, whereas real
operations often use aged accounts. In the persona campaigns, 3--6\% of
narrative posts are verbatim duplicates across channels, where a channel's
rewrite failed and the seed text was used instead; identical-text detection
remains at chance for these campaigns. The cover traffic of campaign channels is
generated from the same day's wire headlines, most of which come from Russian
state outlets, which may make covers less varied than real ones. Detection is
evaluated at the level of channels within a two-week window; other units
(messages, narratives) and longer windows may give different results. Finally,
because campaign text and TeleTwin agents are generated by the same model,
Tier~B tests coordination signals only; it does not test detectors of
machine-generated text.

\section{Ethics}
\label{sec:ethics}

\paragraph{Data.} We collected only public broadcast channels through the
official Telegram API. We stored no information about subscribers or other
individual users, and no private groups or direct messages. All analyses are at
the level of channels, which are public outlets addressing an open audience.
Where we name channels in the paper, they are public media outlets, official
accounts or channels with large public audiences. Released data consist of channel and
message identifiers and derived statistics, in line with the platform's terms,
rather than full message texts~\cite{antonakaki2026teletwinzenodo}.

\paragraph{Synthetic campaigns.} Building a detection benchmark requires
generating campaign content, which carries a risk of misuse. We limit this risk
in four ways. First, every narrative is fictional: it concerns invented
organisations, outlets or vague rumours, never real people, groups or events,
and links point to reserved \texttt{.example} domains. Second, campaign channels
exist only in our data files and in the simulation; nothing was ever posted to
Telegram or any other platform. Third, we will release campaign labels,
detector code, configuration of the design factors and evaluation scripts, but
not the narrative texts or the prompts used to generate campaign content.
Fourth, the strategies we simulate are already documented in public research on
influence operations; the benchmark measures how detectors cope with them and
adds no new capability for attackers.

\paragraph{Simulating real channels.} TeleTwin agents are conditioned on the
public posts of real channels so that the simulated ecosystem resembles the real
one. Simulated posts are never attributed to the real channels outside the
simulation, are not published, and are used only in aggregate. We do not
release simulated text attributed to real channel names.

\paragraph{Dual use and fairness.} Coordination detectors can be used against
legitimate, organically aligned communities. We therefore report false-positive
rates separately for each political perspective, and we caution against
deploying any detector evaluated here without human review.
\section{Conclusion}
We introduced TeleTwin, an LLM-powered digital twin of a Telegram
crisis-discourse ecosystem built from 1.07 million messages of 195 public
channels, and used it to study the detection of LLM-powered information
operations. Evaluated on a held-out period that includes the June 2025
Iran--Israel war, the twin captures the timing of the ecosystem's response to
events and the identity of its hubs better than statistical baselines, but its
agents are more homogeneous than real channels: they over-focus on the
conflict, over-amplify cascades and write recognisably LLM-like text. In real
Telegram data, classic coordination detectors fail against trivial LLM
strategies, and semantic similarity recovers loud campaigns but not quiet ones,
with false positives that fall unevenly across political perspectives. Inside
the twin, detection fails altogether, because the simulated background is about
40 times more semantically homogeneous than the real one, and organic agents
that adopt a campaign's narrative are mistaken for its members. Digital twins
can create the ground truth that real platforms lack, but only if the
diversity of their simulated populations is calibrated as carefully as their
activity. Future work should calibrate homogeneity directly, extend the corpus
to Hebrew and Arabic, use several models and runs, and develop detectors that
use signals beyond content similarity.

\section*{Data and Code Availability}
Code, crawl configuration, channel metadata with perspective labels, message
identifiers and metadata (without message texts), agent profiles (without
style-example texts), campaign ground-truth labels and all derived results are
available on Zenodo~\cite{antonakaki2026teletwinzenodo}:\\
\url{https://zenodo.org/records/23187748}
(DOI \href{https://doi.org/10.5281/zenodo.23187748}{10.5281/zenodo.23187748}).
Message texts can be
re-collected from the public channels with the Telegram API using the released
identifiers. Simulated texts, campaign narratives and the campaign generator
are withheld for the reasons given in Section~\ref{sec:ethics}; the generator is
available to researchers on request.

\section*{Acknowledgments}
This work was supported by the European Union's Horizon Europe research and
innovation programme under grant agreement No.\ 101168438 (INTACT) and grant
agreement No.\ 101249446 (DETANGLE). Views and opinions expressed are those of
the authors only and do not necessarily reflect those of the European Union or
the European Research Executive Agency. Neither the European Union nor the
granting authority can be held responsible for them.

\paragraph{Use of generative AI.} Generative AI tools were used to assist with
code development and with the drafting and editing of the manuscript. Separately,
and as part of the research itself, an open-weights large language model
(Qwen2.5-72B-Instruct) was used to simulate channel agents and to generate the
synthetic campaigns, as described in Sections~\ref{sec:twin}
and~\ref{sec:rq2}. The authors designed the study, verified all code, data and
results, and take full responsibility for the content of this paper. The code
and data supporting the findings are available on
Zenodo~\cite{antonakaki2026teletwinzenodo}.

\bibliographystyle{plain}
\bibliography{references}

\appendix
\section{Simulator Tuning Runs}
\label{app:tuning}
Table~\ref{tab:tuning} lists every tuning run, in order. All windows lie in the
calibration period except the initial smoke test, which only checked that the
pipeline ran and changed no setting. \emph{Sim/real} compares simulated with
real volume of the same agents in the same hours; \emph{sim/avg} compares it
with the average-day expectation implied by the agents' calibration rates.
Gaza/Iran share of simulated messages on the 26 May window (real: 19\%): run~6
(v1) 52\%, run~7 (v2) 40\%, run~8 (v3) 42\%.

\begin{table}[h]
\centering
\small
\caption{Tuning runs (calibration period only).}
\label{tab:tuning}
\setlength{\tabcolsep}{4pt}
\resizebox{\textwidth}{!}{%
\begin{tabular}{lllrrr}
\toprule
Run & Window (2025) & Change & Sim/real & Sim/avg & Forwards (sim / real) \\
\midrule
smoke & 12 Jun 18--21h & first version (pipeline check) & 0.91 & -- & 21\% / 9\% \\
1 & 20 May 12--18h & explicit post and forward counts in prompt & 0.77 & -- & 2\% / 11\% \\
2 & 20 May 12--18h & softer forward wording; wake scale 1.5$\to$1.8 & 0.92 & -- & 24\% / 11\% \\
3 & 20 May 12--18h & forward budget: rate calibrated, choice by LLM & 0.78 & -- & 10\% / 11\% \\
4 & 8 May 06--12h & wake scale 2.3 (new window) & 0.64 & -- & 12\% / 10\% \\
5 & 8 May 06--12h & several sessions per hour ($m=1.9$) & 0.65 & 0.80 & 4\% / 10\% \\
6 & 26 May 00--12h & session scale $\kappa=1.25$; ``forward exactly $b$'' (v1) & 0.93 & 1.01 & 7\% / 9\% \\
7 & 26 May 00--12h & style examples proportional to topic mix (v2) & 0.94 & 1.01 & 8\% / 9\% \\
8 & 26 May 00--12h & v2 + usual Gaza/Iran share in prompt (v3, discarded) & 1.10 & 1.19 & 6\% / 9\% \\
\bottomrule
\end{tabular}}
\end{table}

\section{Detailed Tier A Results}
\label{app:tierA}
\begin{table}[h]
\centering
\small
\caption{RQ2 Tier A: detecting injected campaigns among up to 195 real channels.
ROC-AUC per detector (0.5 = chance); AP and precision@K for the semantic
detector, averaged over three independently generated versions of each campaign
(different cover channels and newly generated text; standard deviations are at
most 0.04 for AUC and 0.12 for precision@K). All campaigns have $K=10$ channels,
staggered or synchronised timing, no links and four narrative posts per channel
per day, unless stated otherwise.}
\label{tab:tierA}
\setlength{\tabcolsep}{4pt}
\resizebox{\textwidth}{!}{%
\begin{tabular}{lrrrrrrr}
\toprule
Campaign & Identical text & Co-URL & Synchrony & Lexical & Semantic & Sem.\ AP & Sem.\ P@K \\
\midrule
Copy, sync & 1.00 & 0.51 & 0.95 & 1.00 & 1.00 & 1.00 & 1.00 \\
Copy, stagger & 1.00 & 0.49 & 0.42 & 1.00 & 1.00 & 1.00 & 1.00 \\
Paraphrase, sync & 0.41 & 0.49 & 0.95 & 0.86 & 1.00 & 1.00 & 1.00 \\
Paraphrase, stagger & 0.41 & 0.49 & 0.43 & 0.88 & 1.00 & 1.00 & 1.00 \\
Cross-lingual, sync & 0.42 & 0.51 & 0.95 & 0.92 & 1.00 & 1.00 & 1.00 \\
Cross-lingual, stagger & 0.44 & 0.49 & 0.49 & 0.91 & 1.00 & 1.00 & 1.00 \\
Persona, sync & 0.46 & 0.51 & 0.95 & 0.91 & 1.00 & 1.00 & 1.00 \\
Persona, stagger & 0.41 & 0.51 & 0.46 & 0.93 & 1.00 & 1.00 & 1.00 \\
\midrule
Paraphrase, stagger, $K=5$ & 0.42 & 0.49 & 0.43 & 0.63 & 1.00 & 1.00 & 1.00 \\
Paraphrase, stagger, $K=20$ & 0.53 & 0.50 & 0.46 & 0.95 & 1.00 & 1.00 & 1.00 \\
Copy, sync, with link & 1.00 & 1.00 & 0.95 & 1.00 & 1.00 & 1.00 & 1.00 \\
Paraphrase, stagger, with link & 0.45 & 1.00 & 0.43 & 0.98 & 1.00 & 1.00 & 1.00 \\
\midrule
Copy, sync, crisis & 1.00 & 0.48 & 0.97 & 1.00 & 1.00 & 1.00 & 1.00 \\
Paraphrase, stagger, crisis & 0.39 & 0.50 & 0.48 & 0.81 & 1.00 & 1.00 & 1.00 \\
Persona, stagger, crisis & 0.36 & 0.48 & 0.44 & 0.84 & 1.00 & 0.95 & 0.87 \\
\midrule
Paraphrase, stagger, generic & 0.43 & 0.49 & 0.46 & 0.74 & 1.00 & 1.00 & 1.00 \\
Cross-lingual, stagger, generic & 0.46 & 0.49 & 0.49 & 0.80 & 1.00 & 1.00 & 1.00 \\
Persona, stagger, generic & 0.43 & 0.49 & 0.42 & 0.92 & 1.00 & 0.98 & 0.90 \\
Paraphrase, stagger, generic, 1/day & 0.42 & 0.51 & 0.44 & 0.75 & 0.94 & 0.32 & 0.10 \\
Persona, stagger, generic, 1/day & 0.42 & 0.49 & 0.54 & 0.87 & 0.87 & 0.18 & 0.03 \\
\bottomrule
\end{tabular}}
\end{table}

\end{document}